\documentclass[letter,10pt]{article}
\usepackage{hyperref}
\usepackage{color}
\usepackage{parskip}
\usepackage[affil-it]{authblk}
\usepackage{graphicx}
\usepackage[title]{appendix}
\usepackage{amssymb}
\usepackage{amsthm}
\usepackage{amsmath}
\usepackage[margin=1in]{geometry}

\usepackage{listings}

\definecolor{dkgreen}{rgb}{0,0.6,0}
\definecolor{gray}{rgb}{0.5,0.5,0.5}
\definecolor{mauve}{rgb}{0.58,0,0.82}
\definecolor{backcolour}{rgb}{0.97,0.97,0.97}
\title{\textsc{XtalOpt} Version 13: Multi-Objective Evolutionary Search for Novel Functional Materials}

\author[a]{Samad Hajinazar}
\author[a,*]{Eva Zurek}
\affil[a]{Department of Chemistry, State University of New York at Buffalo, Buffalo, New York  14260-3000, United States}
\affil[*]{Corresponding author: ezurek@buffalo.edu}
\date{}                     %% if you don't need date to appear
\begin{document}

\maketitle

\begin{abstract}
Version 13 of \textsc{XtalOpt}, an evolutionary algorithm for crystal structure prediction, is now available for download from the CPC program library or the \textsc{XtalOpt} website, \href{https://xtalopt.github.io}{https://xtalopt.github.io}.
In the new version of the \textsc{XtalOpt} code, a general platform for multi-objective global optimization is implemented.
This functionality is designed to facilitate the search for (meta)stable phases of functional materials through minimization of the
enthalpy of a crystalline system coupled with the simultaneous optimization of any desired properties that are specified by the user.
The code is also able to perform a constrained search by filtering the parent pool of structures
based on a user-specified feature, while optimizing multiple objectives.
Here, we present the implementation and various technical details,
and we provide a brief overview of additional improvements that have been introduced in the new version of \textsc{XtalOpt}.
\end{abstract}

%\begin{keyword}
%Neural network interatomic potentials; Evolutionary structure optimization; Molecular dynamics.
%\end{keyword}


\section{Introduction}\label{sec:intro}

The computational prediction of novel materials has recently come to the fore. Prolific examples include GNoME's purported discovery of over 2 million crystalline compounds whose energies fall below the currently-known convex hulls~\cite{GoogUniverse}, and the machine-accelerated high-throughput identification  of ambient-pressure superconductors~\cite{Sanna:2024a,Dolui:2024a}. In this manuscript we describe a general method,  an example of multi-objective optimization, that can be used to predict novel materials with specific functionalities. This method, implemented within the \textsc{XtalOpt} program package,  can be paired with any external optimizer of crystalline lattices, along with any program that can estimate a property of a given material or output a descriptor that serves as a proxy for the desired functionality. Therefore, this new methodology can be used for the discovery of functional materials, but, notably, it can target any property that can be computed through the user's choice of theoretical approach including the predictive models such as those generated in the above examples ~\cite{GoogUniverse,Sanna:2024a,Dolui:2024a} .

In the last two decades, global optimization (GO) algorithms, typically interfaced with density functional theory (DFT) optimizers, have become an integral part of the materials-scientist's toolbox to find approximate solutions for the global minimum in the potential energy surface (PES). The basic techniques include random search~\cite{pickard2009structures}, simulated annealing~\cite{kirkpatrick:1983}, metadynamics~\cite{martovnak2003predicting}, minima~\cite{goedecker2004minima} and basin~\cite{bh0} hopping, particle swarm optimization~\cite{Wang:2010a}, and evolutionary algorithms
\cite{Oganov:2006, Zurek:2011a, SH07, Tipton:2013a}.
Until recently~\cite{Kruglov2023,Salzbrenner2023}, these GO strategies have been used to minimize the 0~K energy or enthalpy of a chemical system, and finite temperature terms have been ignored due to their computational expense. While some of these algorithms are more suited to a global exploration of the PES, others sample local regions and are therefore more likely to find metastable structures. Numerous manuscripts have reviewed these methods, their successes, failures, and more~\cite{Falls2021a,Zurek:2014d,Conway2023,Oganov:2019a,Wang2022}.

Despite the role that GO algorithms have played in aiding the characterization of synthesized materials and predicting novel materials for synthesis~\cite{Zurek:2014i}, their focus on finding the minimum in the PES can be a limitation. Indeed, most of the organic compounds that we know are metastable, and in the solid state various synthesis techniques for accessing metastable phases are actively being developed~\cite{Parija:2018}. The intense interest in metastable compounds stems from their potential uses in a wide variety of technological applications including as superconductors, superhard or refractory materials, thermoelectrics, photovoltaics, and multiferroics. Simply put: GO algorithms may fail to predict a phase with desirable characteristics because it is metastable.

One way that researchers have worked around this limitation is to perform GO searches that locally optimized each structure, but the fitness was calculated using the property of interest rather than the enthalpy. This strategy has been adopted for  the design and prediction of phases with desired electronic structures~\cite{Franceschetti1999, Dudiy2006}, as well as super-dense \cite{zhu2011denser}, super-hard \cite{Lyakhov:2011a} and high dielectric \cite{zeng2014evolutionary,Qu2020} materials. Other techniques that have been suggested include encoding the desired properties into the structural information and evolving them during the search process \cite{Higgins:2019}, or removing structures with undesirable geometric characteristics from the breeding pool by artificially assigning them a high enthalpy~\cite{SH10}.

%Compared to the secondary assessment workflow, this approach is more effective in guiding the search algorithm. Nevertheless, it requires a problem-specific customization in the description of the structural properties and possibly the search algorithm itself.

A more natural integration of the desired properties in the GO process can be achieved by simultaneous optimization of all target objectives. This can be addressed using the family of ``multi-objective" global optimization (MOGO) algorithms \cite{Giagkiozis2015a, Gunantara2018}. Since it is unlikely that the optimal solutions for two (or more) objectives are the same, typical MOGO algorithms aim to find a set of solutions that offer the best trade-off between various objectives. A common approach, the family of Pareto-based algorithms, searches for a set of solution candidates (known as Pareto optimal solutions) such that for any solution no single objective can be improved without worsening the others. Another common technique, belonging to the category of decomposition-based methods, is the scalarizing approach wherein a scalar fitness measure is employed to represent the optimality of the candidate solutions. This approach, effectively, transforms the MOGO problem into a single-objective problem tractable by traditional GO algorithms.

Various MOGO algorithms have been utilized for materials discovery \cite{liao_metaheuristic-based_2020}. Most notable are several studies dedicated to the prediction of new functional materials through implementations of Bayesian global optimization \cite{solomou2018, Gopakumar2018, khatamsaz2022}, and evolutionary and differential evolution algorithms \cite{Yang2008a, chen_predicting_2014, Zhang2015, NUNEZVALDEZ2018152, Maldonis2019, Cheng2020, meng_experimentally_2023}. The focus of this manuscript is on the family of evolutionary search (ES) algorithms, which are iterative stochastic approaches to GO. Among the MOGO approaches, in particular, the multi-objective evolutionary search (MOES) algorithms \cite{Horn1994,srinivas1994a,deb2011abc} are popular because they are inherently population-based and straightforward to implement.

The previous version of the \textsc{XtalOpt} code \cite{ Zurek:2018j, Falls2021a} included an implementation of the MOES algorithm to search for (meta)stable phases of hard materials. In this scheme, the Vickers hardness, obtained from the AFLOW machine learning (ML) estimated shear modulus \cite{Zurek:2019b}, was used in combination with the enthalpy to evaluate the fitness of candidate structures. In addition to finding numerous superhard carbon polymorphs, a superhard-superconducting phase was found~\cite{Wang2022a}.

In this work, we present a general implementation of the \textsc{XtalOpt} MOES algorithm. The new implementation is designed to (i) accommodate an arbitrary number of target objectives, (ii) allow the user to introduce any desired objective as long as it can be represented with a numerical value,  and (iii) offer a general and easy-to-setup interface that can be used with any external code. Further, the \textsc{XtalOpt} MOES can perform various types of optimizations (minimization or maximization) depending on the target objective and can facilitate a constrained search by filtering the pool of structures based on user-specified criteria. This functionality is available in both the command-line interface (CLI) and the graphical user interface (GUI) modes of \textsc{XtalOpt}.

This article describes the MOES functionality recently implemented in \textsc{XtalOpt}. Section \ref{sec:mo-intro} introduces the generalized fitness function: it describes the main input parameters of the MOES and details the properties of the required external codes.  Sections \ref{sec:cli} and \ref{sec:gui} illustrate how to set up a search in the CLI and GUI versions of \textsc{XtalOpt}, respectively, while Section \ref{sec:example-scripts} provides examples of scripts that can calculate objectives for either version. The constrained search functionality is introduced in Section \ref{sec:filtration}, and Section \ref{sec:hardness} describes the changes in the (legacy) hardness optimization. The MOES output and general error handling is outlined in Section \ref{sec:errors-outputs}. Finally, Section \ref{sec:misc} lists a series of options and functionalities, independent of the MOES, that have been added to the latest version of the \textsc{XtalOpt} code.

\newpage
\section{Multi-objective search}\label{sec:mo-intro}

\subsection{Generalized fitness function}\label{subsec:method}

A typical ES \cite{Zurek:2014d} workflow begins with (i) generating a population of candidate structures, (ii) locally relaxing these structures, and (iii) assigning a fitness to a structure based on its enthalpy. From the pool of possible parents (iv) a structure is chosen randomly, but with a probability related to its fitness, to (v) generate a new child structure via applying variations of bio-inspired genetic operators, which include single parent distortions (mutations) or two-parent cut-and-splice (breeding). Steps (ii)-(v) are repeated until a pre-defined stopping criterion is satisfied.

In a single-objective ES, such as the initial implementation of \textsc{XtalOpt}, the fitness, $f_s$, is assigned to each candidate structure $s$ via:

\begin{equation}
f_s=\frac{H_\text{max} - H_{s}}{H_\text{max}-H_\text{min}}
\end{equation}

where $H_{s}$ is the enthalpy of structure $s$, and $H_\text{max}$ and $H_\text{min}$ are the maximum and minimum enthalpies, respectively, of the relaxed structures. The calculated fitness values fall between 0 and 1.0, and the (user defined number of) structures with the highest fitness are selected to be part of the breeding pool. The fitness of the structures comprising the breeding pool is first normalized so its sum is equal to unity, then employed to determine a probability for each structure. The breeding pool candidates are sorted according to the probabilities, normalized such that the lowest enthalpy crystal is assigned a probability of 1.0. A structure is chosen to be a parent if its probability is above that of a random number that falls between 0 and 1. This procedure gives a relatively higher chance for procreation to those candidates that are more ``fit", i.e., have a lower enthalpy.

The MOES, however, needs a different measure for evaluating the suitability of candidate structures to be parents for the next generation. Since an ES is typically followed by high-accuracy local optimization of the best candidates, we found it sufficient to resort to the scalarizing technique by using a generalized scalar function that merges all objectives to obtain a single measure of fitness for each structure~\cite{Emmerich2018}. Assuming that $\{X\}$ and $\{Y\}$ represent sets of objectives to be minimized and maximized, respectively, the generalized fitness for the $s^\textrm{th}$ structure can be obtained as:

\begin{equation}
f_s=
\sum_i w^i_X \left( \frac{X^i_{max} - X^i_{s}  }{X^i_{max}-X^i_{min}} \right) +
\sum_j w^j_Y \left( \frac{Y^j_{s}   - Y^j_{min}}{Y^j_{max}-Y^j_{min}} \right),
\label{eq:fitness}
\end{equation}

where $\{X_s\}$ and $\{Y_s\}$ are the values of the corresponding objectives calculated for the $s^\textrm{th}$ structure, and $\{w\}$ is the set of weights associated with the objectives chosen to reflect their relative significance.

\begin{figure*}
\centering
\includegraphics[width=0.6\linewidth]{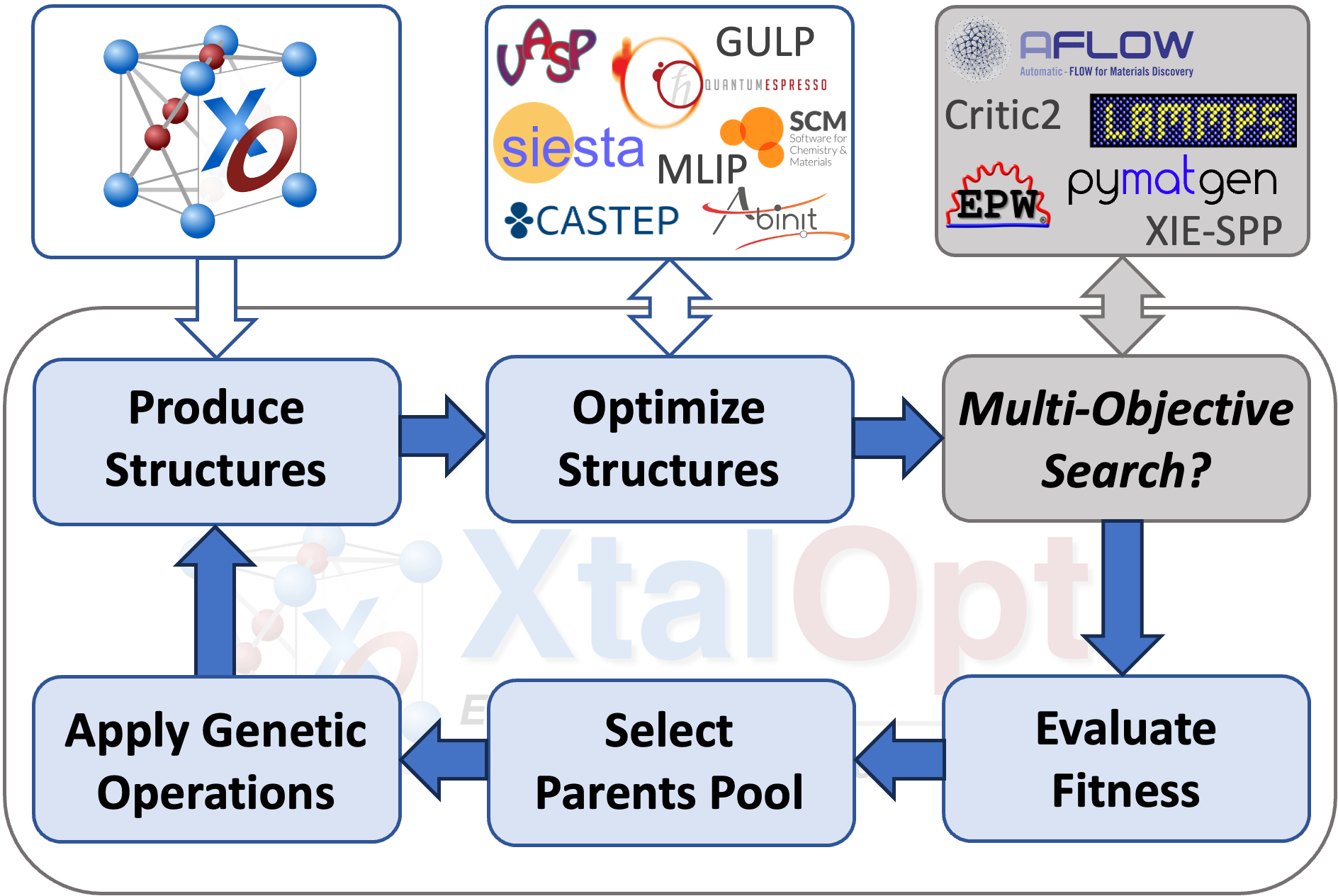}
\caption{Workflow of the \textsc{XtalOpt} MOES (Multi-Objective Evolutionary Search) algorithm. After producing the initial set of structures, local relaxations can be performed via various first-principles approaches or by using interatomic potentials. The locally optimized structures are then passed on to external code(s) introduced by the user for calculating desired properties (a few are shown as examples). Subsequently, the fitness function is evaluated for all structures and the parent pool is selected accordingly. New structures are generated by applying various evolutionary operations to the structures chosen from the parent pool.} \label{fig:0}
\end{figure*}

Given a total weight of $1.0$ for all objectives, the above fitness measure will be normalized to $[0,1.0]$, and the MOES is converted to a single-objective search for which the standard ES workflow can be followed. Figure \ref{fig:0} illustrates schematically the workflow of the MOES implementation within \textsc{XtalOpt}. Following a local optimization, the structural coordinates are used to calculate the desired objectives, whose values are employed to obtain the generalized fitness function for choosing the next parent(s). Otherwise, the workflow resembles that of a single-objective ES.

In practice, enthalpy is always set to be one of the objectives to be minimized. While this is not necessary in our implementation, inclusion of enthalpy enables the search for low lying local minima that potentially feature the target properties, though they may not have the lowest enthalpy. This procedure effectively increases the chance of finding ``metastable" phases with the desired properties. Relying on the target property alone may result in the identification of phases with too-high-enthalpies that can never be synthesized, or those located in extremely shallow potential wells whose barriers can be easily overcome at ambient temperatures.

For utilizing the \textsc{XtalOpt} MOES, the user must specify a (potentially external) code to compute the target property for each objective considered in the search, and provide the corresponding weight for the fitness calculation (Equation \ref{eq:fitness}), as described in the next section.

\subsection{User-defined objectives}\label{subsec:user-defined-objectives}

For any property employed as an objective in the MOES, there should be an (executable) user-provided script or code that (i) calculates that property from the coordinates of a structure generated during the course of a regular \textsc{XtalOpt} search, and (ii) produces an output file with a single number (integer or double) as the value of the desired objective. The script should be available either in the local path (for a local run) or on the cluster access path (for a remote queue). \textsc{XtalOpt} will call this script automatically and will read and use its output for the MOES; either for determining a structure's fitness for procreation or for filtering the parent pool (see Section \ref{sec:filtration}).

After each local relaxation is performed using any of the total energy calculation methods available in \textsc{XtalOpt}, a structure file is generated. This file is named \texttt{output.POSCAR}, irrespective of the external optimizer, and it is written in the format used by the Vienna Ab initio Simulation Package (VASP) \cite{VASP1,VASP2}. The user-provided script should be able to read and employ this file (or convert it to another structural data format if needed) to perform the intended calculations and produce the required output file, which should be a text file with the numerical value of the corresponding objective written as the first entry of the first line of the file.

The aforementioned script can simply contain a sequence of commands that use the \texttt{output.POSCAR} file and call some program to produce a result. However, if the intended calculations are computationally demanding, it may be preferable if the script generates a cluster job file and submits the job to the computational cluster. If input data not present in the output of the structural relaxation is required by the script (e.g., a file containing the \textit{ab initio} charge density, the density of states at the Fermi level, etc.), commands to generate this data should be present in the script. Alternatively, the required values or files can be obtained by adding the appropriate entries to the \textsc{XtalOpt} job template used in the structure search. Samples scripts are provided in Section \ref{sec:example-scripts}.

\subsection{Multi-objective search parameters}\label{subsec:mo-search-parameters}

To invoke the MOES functionality the user should provide the following information to the \textsc{XtalOpt} code for every desired objective:

\begin{itemize}
\item \textbf{optimization type}: instructions on how to use the value of a calculated objective in determining a structure's fitness. \textsc{XtalOpt} can minimize or maximize an objective, filter the parent pool according to user-defined criteria (see Section \ref{sec:filtration}), or maximize the AFLOW-ML hardness (see Section \ref{sec:hardness}).

\item \textbf{path to the user-defined script}: the full path to the script that retrieves or calculates the desired property corresponding to the introduced objective. The script is automatically run by the \textsc{XtalOpt} code for each locally relaxed structure.

\item \textbf{script's output filename}: the name of the file generated by the script that contains the calculated result of the corresponding objective.

\item \textbf{optimization weight}: a number between 0.0 and 1.0 used as the weight for the corresponding objective in the fitness calculation. The total weight of all objectives should not exceed 1.0, and the weight for minimizing the enthalpy is calculated by \textsc{XtalOpt} as: ``\textit{1.0 - total weight of the objectives}".
\end{itemize}

Any objective explicitly assigned a weight of 0.0 will be calculated, but will not affect the optimization. Moreover, if the sum of the weights explicitly assigned equals 1.0, the enthalpy is assigned a weight of 0.0, i.e., the fitness is determined only based on the values calculated for the other objectives. If the total weight of the introduced objectives exceeds 1.0, \textsc{XtalOpt} will quit after producing an error message.

\section{Multi-objective run in the \textsc{XtalOpt} CLI}\label{sec:cli}

In the MOES run in the CLI mode, for each user-defined objective a line should be added to the \textsc{XtalOpt} input file that starts with the keyword \textit{objective}. This line includes the above-mentioned information for the objective and, generally, has the following format:

{\begin{lstlisting}[language=bash]
objective = "optimization_type" "/path/script" "script_output_filename" "weight"
\end{lstlisting}}

It should be noted that in the CLI mode of \textsc{XtalOpt}:

\begin{enumerate}

\item Possible options for the ``\textbf{optimization\_type}" are ``minimization", ``maximization", ``hardness", and ``filtration", as introduced above. This field is not case-sensitive, and only the first three letters are important in identifying the optimization type by \textsc{XtalOpt} (i.e., ``min", ``max", ``har", and ``fil").

\item Providing the ``\textbf{script\_output\_filename}" is \textit{optional}. If this is not specified, the default will be \texttt{objective\#.out} in which ``\texttt{\#}" is the number of the objectives in the order that they appear in the \textsc{XtalOpt} input file (excluding the ``hardness" objective), e.g., \texttt{objective1.out}, \texttt{objective2.out}, etc.

\item Specifying the ``\textbf{weight}" for the objective is \textit{optional}, as well. If this field is not given for a number of objectives, it will be calculated. Specifically, if any weight is provided for any of the objectives, it will be subtracted from 1.0 and the remaining value will be divided between the enthalpy and objectives that don't have a specified weight.
\end{enumerate}

%\subsection{Examples of \textsc{XtalOpt} input file entries}\label{subsec:input-file-entries}

%\comment{Just a point to consider: the introductory paragraph here, technically speaking, seems to be more relevant to the "example scripts" which is the next section. The reason that I separated that (i.e., not as a subsection like this one), is that the scripts apply to both the CLI and GUI, whereas the input parameters here are relevant to the CLI case only.}

Let us now provide some examples of acceptable input files. For a calculation that aims to minimize the volume per atom and maximize the electronic band-gap, the following lines can be added to the \textsc{XtalOpt} input file:

{\begin{lstlisting}[language=bash]
objective = min    /path/vol.sh    vol.dat    0.2
objective = max    /path/gap.sh    gap.dat    0.2
\end{lstlisting}}

In this case, \texttt{vol.sh} and \texttt{gap.sh} are two executable scripts (either in a local or remote location, depending on the run) that use the \texttt{output.POSCAR} file to calculate the volume per atom and the band-gap, respectively. \textsc{XtalOpt} expects the calculated values to be written by these scripts to the \texttt{vol.dat} and \texttt{gap.dat} files in the structure's directory. Since the weight for both objectives is 0.2, the remaining weight of 0.6 will be assigned to the enthalpy.

The following example illustrates the flexibility of the input entries:

{\begin{lstlisting}[language=bash]
objective = min    /path/vol.sh    vol.dat
objective = max    /path/gap.sh    0.2
\end{lstlisting}}

Since the weight for the volume objective is not specified, the remaining total weight of 0.8 will be divided equally between enthalpy and volume per atom, and since the output filename for the band-gap calculation is not given, \textsc{XtalOpt} will expect a file named \texttt{objective2.out} (as this is the second objective in the list of objectives) to be present with the corresponding value.

\section{Multi-objective runs in the \textsc{XtalOpt} GUI}\label{sec:gui}

\begin{figure*}
  \centering
  \includegraphics[width=\linewidth]{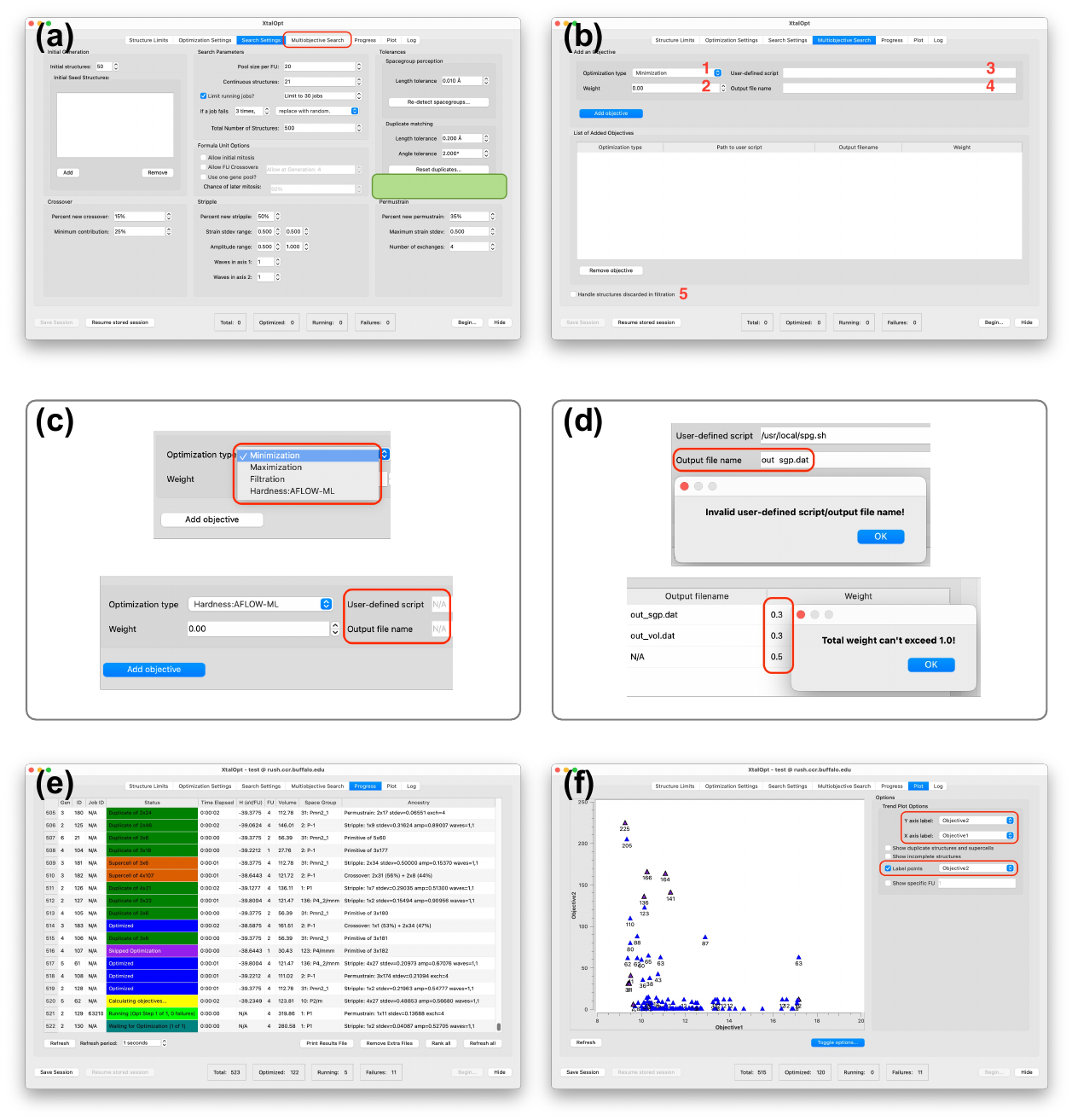}
  \caption{Screenshots from a MOES run in the \textsc{XtalOpt} GUI mode.
        (a) The \textbf{Search Settings} tab no longer includes AFLOW-ML hardness entries (green  oval),
          and a new \textbf{Multiobjective Search} tab (b) is added.
        (c) Drop-down menu showing the multi-objective search type options, including for an AFLOW-ML hardness calculation.
        (d) Pop-ups showing errors that arise from non-acceptable text entries, and entering weights that sum to a value larger than one.
        (e) The \textbf{Progress} tab with new MOES-related status.
        (f) The \textbf{Plot} tab in a MOES run with user-defined objectives as labels and tags.} \label{fig:1}
\end{figure*}

We now describe how the \textsc{XtalOpt} GUI has been modified to reflect the new options that are available in the MOES run. In the \textbf{Search Settings} tab the AFLOW-ML hardness related entries have been removed (green box in Figure \ref{fig:1}(a)), and a new \textbf{Multiobjective Search} tab is introduced (red outline). In the \textbf{Multiobjective Search} tab (Figure \ref{fig:1}(b)), all entries relevant to a MOES run (described in Section \ref{subsec:mo-search-parameters}) can be entered in the corresponding fields by (1) choosing the MOES run type from a drop-down menu, (2) specifying the weight associated with a particular objective, entering the (3) name and full path to the user-provided script and (4) output file names, and (5) specifying how \textsc{XtalOpt} should handle a structure that fails a filtration objective (as discussed in Section \ref{sec:filtration}). It should be noted that any weight left as zero in the GUI input will result in the corresponding objective being calculated without affecting the optimization. Finally, clicking the button ``Add objective'' adds the defined objective to the list of objectives for the run.

The aforementioned MOES run specifications in the CLI mode apply to the GUI case, as well:

\begin{itemize}
\item The types of MOES runs are ``maximization", ``minimization", ``filtration", and ``hardness"; and setting up an AFLOW-ML hardness calculation only requires its weight to be specified (Figure \ref{fig:1}(c)),

\item No more than one instance of ``hardness" objective is considered in each run; if more than one is present, only the corresponding weight (if entered differently) will be updated,

\item Desired objectives can be arbitrarily added or removed before the run is started. However, once the search begins, the only MOES-related parameter that can be altered is the one instructing the code how to handle the structures discarded by a ``filtration" objective (as detailed in Section \ref{sec:filtration}).
\end{itemize}

Further, in the \textsc{XtalOpt} GUI MOES implementation,

\begin{itemize}
\item Common errors in the input parameters (e.g., leaving script or file name fields empty, spaces in text entries, total weight exceeding 1.0), result in an error message from the code (Figure \ref{fig:1}(d)),

\item In the \textbf{Progress} tab, the status of a structure that has successfully finished local relaxations changes to ``Calculating objectives...", after which the status changes to ``Optimized", ``ObjectiveDismiss", or ``ObjectiveFail" according to the calculation results (Figure \ref{fig:1}(e)),

\item In the \textbf{Plot} tab, once a MOES run is started, the list of introduced objectives appears among the available options for the $x$ and $y$ axes of the trend plots, as well as the list of labeling symbols (Figure \ref{fig:1}(f)).

\end{itemize}

\section{Examples of user-defined scripts}\label{sec:example-scripts}

In this section we provide an example of a MOES-compatible script that can be employed with both the CLI and GUI versions. We choose a simple example, wherein the goal is to minimize the enthalpy, while simultaneously maximizing a structure's space group number calculated from the VASP format \texttt{output.POSCAR} structure file. This can be done via a simple Python script, e.g., \texttt{/path/spg.py}, where the Atomic Simulation Environment (ASE) \cite{HjorthLarsen2017a} is utilized to resolve the space group of the structure as,

\begin{lstlisting}[language=python]
import ase.io as io
from ase import Atoms
from ase.spacegroup import get_spacegroup
s=io.read('output.POSCAR', index ='-1', format='vasp')
print(get_spacegroup(s, symprec=1e-3).no)
\end{lstlisting}

The output of this short Python code can be used via a simple executable bash script, e.g., \texttt{/path/spg.sh},

\begin{lstlisting}[language=bash]
#!/bin/bash
/path/python /path/spg.py > spg.dat
\end{lstlisting}

to produce a \texttt{spg.dat} file containing the space group number of the structure, which is readable by \textsc{XtalOpt} for this objective. In the \textsc{XtalOpt} input file for the CLI mode, this can be then introduced as:

\begin{lstlisting}[language=bash]
objective = max    /path/spg.sh    spg.dat
\end{lstlisting}

along with the desired weight (or leaving the weight un-specified for \textsc{XtalOpt} to adjust it).

Alternatively, if the user desires to submit this calculation to a computational cluster, the executable script \texttt{/path/spg\_queue.sh} in its most basic form can be written as:

\begin{lstlisting}[language=bash]
#!/bin/bash

cat > fspg.slurm << EOF
#!/bin/bash
#SBATCH --nodes=1 --ntasks-per-node=1
#SBATCH --job-name=fspg
#SBATCH --output=fspg.out --error=fspg.err
#SBATCH --time=00:05:00
#SBATCH --cluster=slurm

##===== main task: calculating the objective
/path/python /paht/spg.py > spg.dat
##=====

EOF

sbatch fspg.slurm
\end{lstlisting}

This particular executable script writes a job submission script for the slurm cluster to disk (\texttt{fspg.slurm} file, which includes everything between the lines containing the ``\texttt{EOF}" keywords) and then submits this job (with ``\texttt{sbatch fspg.slurm}") to the cluster. The job submission script (\texttt{fspg.slurm} file) includes an introductory part (job- and cluster-related settings) and the core task, just like a usual job submission script. It contains the previous simple script for calculating the space group number (which is enclosed between ``\texttt{\#\#=====}" comment lines for clarity). The latter script can be employed in conjunction with the MOES to optimize the space group number just as in the previous example, only, this time each calculation is submitted to the cluster instead of running through a simple executable bash script.

While the examples provided in Section \ref{sec:cli} and Section \ref{sec:example-scripts} illustrate the structure of the input file that must be present when the CLI version is employed and the files required to calculate the property of interest, the objectives chosen are somewhat artificial. Objectives such as the calculated density of states at the Fermi level, band-gap, Vickers or Knoops Hardness, superconducting critical temperature, $zT$ figure of merit and more could be chosen, as desired by the user. A follow up publication will focus on application of the MOES-\textsc{XtalOpt} methodology for identifying structures with desired properties.

\section{Filtering structures: Constrained search}\label{sec:filtration}

The MOES implementation within the \textsc{XtalOpt} code can also facilitate a \textit{constrained ES}. Besides maximizing or minimizing a particular objective, \textsc{XtalOpt} can optionally filter the relaxed structures based on user defined properties. The constrained search prevents crystals deemed unsuitable from entering the parent pool, hence promoting or prohibiting the propagation of a specific genetic characteristic. A similar approach was employed in Ref.\ \cite{SH10}, except that structures were excluded from the parent pool by artificially setting their enthalpy to an unphysically high value. The filtration technique introduced here has a similar effect, but without the need of modifying the enthalpy.

To utilize the filtration functionality, similar to the ``minimization" and ``maximization" features, the user should provide a script that marks the structures to keep or discard based on the intended property. Structures that are marked for discarding will not be allowed in the parents' pool, although they will remain in the set of generated structures. An example of the relevant entry in the \textsc{XtalOpt} input file for the CLI mode is:

{\begin{lstlisting}[language=bash]
objective = fil    /path/script    out_file
\end{lstlisting}}

The workflow differs compared to when an objective is minimized or maximized by \textsc{XtalOpt} in the following ways. Since the objective is not meant for optimization, regardless of the user-defined weight, the objective's weight will be automatically set to 0.0. Moreover, the numerical value written to the output file by the script should be either 0 (instructing \textsc{XtalOpt} to discard the structure) or 1 (instructing \textsc{XtalOpt} to keep it). A structure that fails the filtration step, by default, will be removed from the parents' pool. However, the user can optionally instruct \textsc{XtalOpt} for further handling of a dismissed structure by adding the text

{\begin{lstlisting}[language=bash]
objectivesReDo = true          # default is false
\end{lstlisting}}

to the input file in the CLI mode. In this case, \textsc{XtalOpt} proceeds depending on the value of the \textit{jobFailAction} flag in the input file. If the \textit{jobFailAction} is set to ``replaceWithRandom" (default value) or ``replaceWithOffspring'' the failed structure is replaced with a new structure generated randomly or by applying evolutionary operations, respectively. On the other hand, if \textit{jobFailAction} is set to ``kill" or ``keepTrying", no further action is taken. If the user instructions result in replacing the failed structure with a new one, it will then be submitted for local optimization and the subsequent calculation of objectives, including the filtration objective. It should be noted that this procedure will be performed at most once for a failed structure, i.e., no more than one replacement will be attempted for a structure that is marked to be dismissed in filtration.

The same can be achieved in the GUI mode by checking the ``Handle structures discarded in filtration" box in the \textbf{Multiobjective Search} tab; where the appropriate follow-up action will be taken according to the ``If a job fails" entry in the \textbf{Search Settings} tab and similar to the above workflow discussed for the CLI case.

\section{AFLOW-ML hardness}\label{sec:hardness}

Previously, maximizing the AFLOW-ML hardness was invoked with the \textit{calculateHardness} flag in the CLI  \textsc{XtalOpt} version (or, setting relevant entries in the GUI). In the current version the Vickers hardness is obtained by introducing the ``hardness" objective. In the CLI mode, this can be performed by adding an objective with the ``hardness" optimization type and, optionally, providing a corresponding weight, i.e.,

{\begin{lstlisting}[language=bash]
objective = hardness   "weight"
\end{lstlisting}}

Therefore, AFLOW-ML hardness calculations are now treated as a user-defined objective. It should be noted that the script name and the output file name inputs do not need to be provided for a ``hardness" objective, and while an arbitrary number of objectives with ``maximization", ``minimization", and ``filtration" type can be introduced, no more than one ``hardness" objective can be added by the user. If there is more than one entry for hardness calculations in the \textsc{XtalOpt} input file, only the last one will be considered by the code (i.e., the weight for hardness calculation will be that of the last entry).

An objective of the ``hardness" type performs the hardness optimization by obtaining the relevant data from AFLOW-ML through internal functions of the \textsc{XtalOpt} code. This is a legacy code and will be disabled in future releases of \textsc{XtalOpt} to avoid compatibility issues. Instead of using this type of objective,  a script to facilitate AFLOW-ML hardness optimization, similar to any regular ``maximization" objective, is strongly encouraged. In this case, an example of a script that can be used to retrieve the required information from the AFLOW-ML platform is presented in \ref{sec:app1}.

\section{Error handling and outputs}\label{sec:errors-outputs}

The external script (or the code it is called by) may fail to operate properly or fail to produce the output in a format that is readable by \textsc{XtalOpt}. In general, \textsc{XtalOpt} considers the output file to be correct and contain a valid value only if the ``first entry" in the ``first line" of the file is a numerical value.  Otherwise, e.g., the file is not produced, is empty, or its first entry is not a legitimate numerical value, the calculation will be marked as failed, and the structure will not be considered as a candidate to enter the breeding pool.

Generally, the search settings (including the MOES settings) are written to the \texttt{xtaloptSettings.log} and \texttt{xtalopt.state} files, which can be used to verify the initialization of the search. In the CLI mode the \texttt{xtalopt-runtime-options.txt} file, which includes the parameters that can be modified during the search, is also produced. Among the MOES-related entries only the \textit{objectivesReDo} flag is output to this file, and is allowed to be changed once the run is started.

For each structure, besides the corresponding output files generated by the scripts, a summary of the objective-related info (overall status of its calculations and their value) is written to the \texttt{structure.state} file.

Finally, just as with any \textsc{XtalOpt} run, the live status is available in the \texttt{results.txt} file, which summarizes the ranking of structures generated during the course of the ES. In the case of MOES runs, the status of a structure that has successfully been locally optimized changes to ``ObjectiveCalculation". Once the objectives are obtained, the status changes to ``Optimized", ``ObjectiveDismiss", or ``ObjectiveFail" depending on whether the calculations finished successfully, the structure was discarded during filtration, or the calculations failed, respectively. Moreover, the \texttt{results.txt} file contains an extra column for the calculated values of each objective introduced by the user.

\section{Miscellaneous}\label{sec:misc}

In the new release of the \textsc{XtalOpt} code, a number of options are implemented to address special situations the user might encounter. These options are briefly introduced in the following.

\subsection{Scaled volume limits (CLI and GUI mode)}\label{subsec:scaled-volume}

An important step in conducting a successful ES is the specification of reasonable minimum and maximum volumes for the unit cells. The previous versions of \textsc{XtalOpt} allowed the user to either explicitly specify these limits, or to introduce a fixed value for the generated unit cells' volume. Now, a new option in \textsc{XtalOpt} can facilitate a reasonable guess.

In the CLI mode, the user can optionally specify the pair of flags:

\begin{lstlisting}[language=bash]
volumeScaleMin = ####
volumeScaleMax = ####
\end{lstlisting}

with the corresponding values being real numbers greater than zero (e.g., 0.8 and 1.2 for the minimum and maximum values, respectively). If these flags are properly provided, \textsc{XtalOpt} first calculates the total volume of spheres of van der Waals radii for all atoms in the formula unit. Then, it multiplies that total volume by the scaling factors to obtain the minimum and maximum volumes. One can check the final calculated values in the run output (i.e., the \texttt{xtalopt.state} or \texttt{xtalopt-runtime-options.txt} files).

\begin{figure}
\centering
\includegraphics[width=0.59\linewidth]{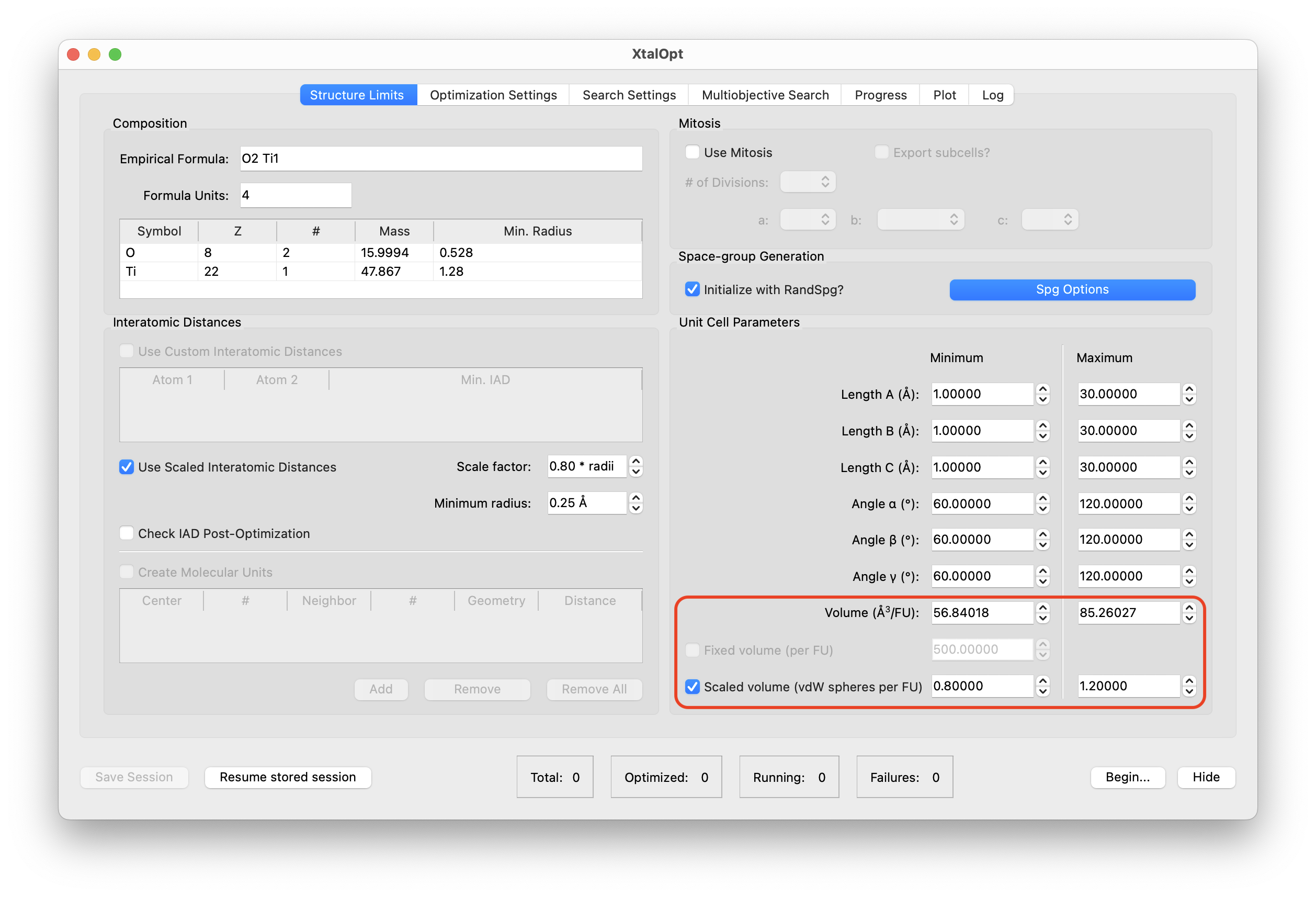}
\caption{Screenshot of the ``Structure Limits'' tab in the new \textsc{XtalOpt} GUI. The red circle highlights the settings used calculated the acceptable volume range from the sum of van der Waals spheres multiplied by a scaling factor.} \label{fig:2}
\end{figure}

The GUI mode includes this option in the \textbf{Structure Limits} tab, where the user can set the scaling factors and access a live update of calculated minimum and maximum volume limits while adjusting the scaling factors (Figure \ref{fig:2}).

\subsection{Running \textsc{XtalOpt} locally on a cluster (CLI mode)}\label{subsec:local-queue}

Often when lengthy ES runs are performed, the \textsc{XtalOpt} code is executed on the cluster where the jobs are being submitted (i.e., running \textsc{XtalOpt} locally while submitting the jobs to a queue). As the jobs are being submitted to the cluster, a remote queue interface should be specified in the \textsc{XtalOpt} input (e.g., slurm, pbs, etc.). This, however, requires a ssh connection to the cluster itself, which might or might not be allowed, depending on the ssh configuration of the user's account. For these types of \textsc{XtalOpt} runs, in the CLI mode, the user can add the following flag to the input file and run the code as a regular remote run:

\begin{lstlisting}[language=bash]
localQueue = true        # default is false
\end{lstlisting}

\subsection{Termination of the \textsc{XtalOpt} run (CLI mode)}\label{subsec:termination}

In a regular \textsc{XtalOpt} run, and once the maximum number of structures (specified by the user) is generated, the user can resume the run by increasing this maximum number (among other run-time flags that can be changed). This functionality requires the code to not quit automatically, and the user needs to terminate the application manually after the desired output is obtained. There are, however, situations where it is desirable for the code to exit after producing a specified number of structures (e.g., scripting a series of \textsc{XtalOpt} runs over multiple directories, such that each run should begin after the previous one is finished). For a run in the CLI mode, the code can be instructed to exit after producing the specified number of structures by adding the following flag to the input file:

\begin{lstlisting}[language=bash]
softExit = true         # default is false
\end{lstlisting}

or by setting its value to true in the run-time setting file \texttt{xtalopt-runtime-options.txt} during the run. With this flag set to true, the code quits after all running (and pending) jobs are finished and the output files are updated.

On the other hand, and at any moment during a run, the user can force the \textsc{XtalOpt} process (hence, the run) to quit immediately by adding the following line to the run-time settings file (\texttt{xtalopt-runtime-options.txt}),

\begin{lstlisting}[language=bash]
hardExit = true
\end{lstlisting}

It should be noted that the \textit{hardExit} flag terminates the \textsc{XtalOpt} running process regardless of any running or pending jobs and without updating the output files, and  this is only a run-time option and the presence of the flag in the input file is ignored by \textsc{XtalOpt}.

\subsection{New options for the VASP optimizer (CLI and GUI modes)}\label{subsec:vasp-new}

Recent versions of the VASP code allow for training and using ML interatomic potentials.
The new version of \textsc{XtalOpt} supports the output (i.e., \texttt{OUTCAR} file) generated by structural optimizations performed using VASP ML force fields.

In previous releases of \textsc{XtalOpt}, when the VASP optimizer was used, a  \texttt{POTCAR} file was required for each element type present in the system. In the new version, it is possible to provide only a single \texttt{POTCAR} file for a multi-element system. This option is especially useful when \textsc{XtalOpt} is interfaced with an external code that is not explicitly supported (e.g., an arbitrary optimizer, which is scripted to produce VASP format output files).

In the GUI, introducing a single \texttt{POTCAR} file for the system can be accomplished by setting the path as:

\begin{lstlisting}[language=bash]
%fileContents:/path/to/system/potcar%
\end{lstlisting}

and in the CLI mode this can be done by introducing a \texttt{POTCAR} file of the ``system" type in the input file, i.e.,

\begin{lstlisting}[language=bash]
potcarFile system = /path_to/potcar
\end{lstlisting}

It should be noted that as \textsc{XtalOpt} arranges the chemical elements in alphabetical order, individual POTCAR files should be combined in the same order to produce the correct results. Moreover, if a ``system" POTCAR is introduced, other entries of \textit{potcarFile} flag in the \textsc{XtalOpt} input file will be ignored by the code.

\subsection{Run-time log file and debug options (CLI and GUI modes)}\label{subsec:debug-log}

In the previous versions of \textsc{XtalOpt}, the code output a comprehensive list of messages regarding the progress of the run in the CLI mode, while only a subset of this information (i.e., key updates about the status of the run) were available in the \textbf{Log} tab for a run in the GUI mode. In the new \textsc{XtalOpt} release,  compilation with the

\begin{lstlisting}
-DXTALOPT_DEBUG=ON
\end{lstlisting}

configuration option saves a detailed list of output messages to the log file \texttt{xtaloptDebug.log} in the local working directory for both the CLI and GUI modes.

Further, if the code is compiled with the

\begin{lstlisting}
-DMOES_DEBUG=ON
\end{lstlisting}

configuration option, running a MOES produces a set of output messages regarding the calculation of objectives and the generalized fitness function. These lines, starting with the keyword \texttt{NOTE}, contain information useful in monitoring the sanity of the calculations and MOES run.

\section{Conclusions}
Herein, we describe developments to the latest version of the \textsc{XtalOpt} evolutionary algorithm that make it possible to perform a multi-objective global optimization (MOGO) search for materials with a desired functionality. The resulting multi-objective evolutionary search (MOES) employs a weighted linear sum of functions for each introduced objective, and it belongs to the category of decomposition-based MOGO techniques. This makes it possible for the user to (optionally) minimize the energy or enthalpy of a crystalline lattice, while at the same time optimizing (minimizing or maximizing) any arbitrary feature or objective for which a numerical value can be obtained, optionally by a code other than the one employed for structural relaxation. A further option to filter structures with undesirable characteristics from the breeding pool is also implemented.

The MOES implementations in both the graphical user interface (GUI) and command line interface (CLI) versions are described, and example input files for the latter are provided. Examples of user defined scripts, which can be used to call the external codes either locally or using a job submission script that is sent to the computational cluster, are also given. Finally, a number of miscellaneous fixes, regarding options for: choosing the unit cell volumes, CLI-mode initialization and termination, run-time and log file generation, and for the VASP optimizer are described.

The MOES implemented in \textsc{XtalOpt} will be useful for the computational discovery of novel materials with a wide range of functionalities. Work is underway in developing descriptors, workflows and user-defined scripts for the prediction of superconductors, electrides, crystalline lattices with user-defined geometric features and more.

\section*{Acknowledgement}
We acknowledge the U.S. National Science Foundation
(DMR-2136038 and DMR-2119065) for financial support, and the
Center for Computational Research at SUNY Buffalo for computational support

(http://hdl.handle.net/10477/79221).

\vspace{\baselineskip}

%\appendix

\begin{appendices}
\renewcommand{\thetable}{\Alph{section}\arabic{table}}

\section{Retrieving AFLOW-ML data}\label{sec:app1}

The data within the AFLOW database has been used to train a number of ML models, for example for obtaining structure-dependent band-gaps (or metal/insulator classification), bulk and shear moduli, (constant volume or pressure) heat capacity, Debye temperature, thermal expansion coefficient, and unit cell energy~\cite{Isayev:2017a}. Given a \texttt{POSCAR} with the VASP format, the following script \cite{Zurek:2017n} can be used to obtain the AFLOW-ML data, to be used in a MOES for one of the aforementioned objectives

\begin{lstlisting}[language=python]
#!/usr/bin/python3
import json, sys, os
from time import sleep
from urllib.parse import urlencode
from urllib.request import urlopen
from urllib.request import Request
from urllib.error import HTTPError
SERVER="http://aflow.org"
API="/API/aflow-ml"
MODEL="plmf"
poscar=open('POSCAR', 'r').read()
encoded_data = urlencode({'file': poscar,}).encode('utf-8')
url = SERVER + API + "/" + MODEL + "/prediction"
request_task = Request(url, encoded_data)
task = urlopen(request_task).read()
task_json = json.loads(task.decode('utf-8'))
results_endpoint = task_json["results_endpoint"]
results_url = SERVER + API + results_endpoint
incomplete = True
while incomplete:
  request_results = Request(results_url)
  results = urlopen(request_results).read()
  results_json = json.loads(results)
  if results_json["status"] == 'PENDING':
    sleep(10)
    continue
  elif results_json["status"] == 'STARTED':
    sleep(10)
    continue
  elif results_json["status"] == 'FAILURE':
    print("Error: prediction failure")
    incomplete = False
  elif results_json["status"] == 'SUCCESS':
    print("Successful prediction")
    print(results_json)
    incomplete = False
\end{lstlisting}

It should be noted that AFLOW-ML models are also available for a series of chemical-formula-only-dependent properties, e.g., vibrational free energies and entropies~\cite{Legrain2017} and the superconducting critical temperatures~\cite{Stanev2018}. However, the outputs of these models are not relevant to the MOES implementation within \textsc{XtalOpt}, which uses a fixed chemical composition during the run.

\end{appendices}

%% References







% \end{thebibliography}

\end{document}